\documentclass[%
aip,
amsmath,amssymb,
reprint,%
]{revtex4-1}
\usepackage{comment}
\usepackage{graphicx}
\usepackage{dcolumn}
\usepackage{bm}
\usepackage{subfig}

\usepackage[utf8]{inputenc}
\usepackage[T1]{fontenc}
\usepackage{mathptmx}
\usepackage{etoolbox}

\makeatletter
\def\@email#1#2{%
	\endgroup
	\patchcmd{\titleblock@produce}
	{\frontmatter@RRAPformat}
	{\frontmatter@RRAPformat{\produce@RRAP{*#1\href{mailto:#2}{#2}}}\frontmatter@RRAPformat}
	{}{}
}%
\makeatother

\begin{document}
	
	\preprint{AIP/123-QED}
	
	\title[Data-Driven Insights into Jet Turbulence: Explainable AI Approaches]{Data-Driven Insights into Jet Turbulence: Explainable AI Approaches}

	\author{E. Amico}
	\affiliation{ Department of Mechanical and Aerospace Engineering, Politecnico di Torino, Torino, Italy}
    	\email{enrico.amico@polito.it}

	\author{L. Matteucci}%
	\affiliation{ Department of Mechanical and Aerospace Engineering, Politecnico di Torino, Torino, Italy}%

    \author{G. Cafiero}%
	\affiliation{ Department of Mechanical and Aerospace Engineering, Politecnico di Torino, Torino, Italy}%

	\date{\today}
	
	\begin{abstract}
In this study, eXplainable Artificial Intelligence (XAI) methods are applied to analyze flow fields obtained through PIV measurements of an axisymmetric turbulent jet. A convolutional neural network (U-Net) was trained to predict velocity fields at subsequent time steps. Three XAI methods—SHapley Additive explanations (SHAP), Gradient-SHAP, and Grad-CAM—were employed to identify the flow field regions relevant for prediction. SHAP requires predefined segmentation of the flow field into relevant regions, while Gradient-SHAP and Grad-CAM avoid this bias by generating gradient-based heatmaps. The results show that the most relevant structures do not necessarily coincide with regions of maximum vorticity but rather with those exhibiting moderate vorticity, highlighting the critical role of these regions in energy transfer and jet dynamics. Additionally, structures with high turbulent dissipation values are identified as the most significant. Gradient-SHAP and Grad-CAM methods reveal a uniform spatial distribution of relevant regions, emphasizing the contribution of nearly circular structures to turbulent mixing. This study advances the understanding of turbulent dynamics through XAI tools, providing an innovative approach to correlate machine learning models with physical phenomena.
    
	\end{abstract}
	
	\maketitle

	\section{Introduction}
	
In recent years, there has been a steady increase in computational power, along with the development of GPUs. These advancements have enabled much faster solutions to numerical problems. In recent literature, it is quite common to find studies where direct numerical simulations (DNS) are performed using GPUs, resulting in significantly shorter computation times compared to traditional CPU-based methods.
	
At the same time, the past decade has seen the rise of machine learning and the development of data-driven techniques, also benefiting from the progress of increasingly powerful and efficient GPUs. In various studies, data obtained from DNS have been used to train neural networks (NN) for a range of applications: prediction, reconstruction, data augmentation, simulation acceleration, and data estimation.
	
	While machine learning techniques offer excellent tools for handling large datasets, it is important to consider that they often function as black-box models. As a result, understanding the internal mechanisms that lead to the results obtained after training can be highly complex, and sometimes impossible. One of the key strengths of neural networks lies in their ability to model highly nonlinear phenomena, which is a central focus of ongoing research in the field.
	
Fluid dynamics is governed by a complex and highly nonlinear system of equations known as the Navier-Stokes equations (NS). Due to this nonlinearity, neural networks represent a valid alternative for modeling the NS. However, it is reasonable to assume that during the training phase with fluid dynamics data—whether numerical (DNS) or experimental—a neural network learns to capture these nonlinearities and, consequently, the underlying physical mechanisms of the phenomenon.
	
The main challenge is that turbulence is a multi-scale phenomenon in both time and space. Energy is primarily transferred from the largest scales to the smallest, where it dissipates, although there is also a reverse energy transfer. In any engineering flow, the difference in magnitude between these scales can be vast, and this scale separation increases with higher Reynolds numbers. The presence of a wall further complicates this energy cascade due to the transfer of energy and momentum from the wall to the external flow. 
    
Jets are characterized by a region of high shear at the jet boundary, where the jet fluid mixes with the surrounding fluid or in the self-similar region. This shear region is unstable and leads to the development of turbulence (coherent structures, ring vortices \citealp{Crow_Champagne_1971,Davies_Fisher_Barratt_1963}). Turbulence in a jet is marked by the presence of vortices of various sizes and time scales. These vortices interact with each other, resulting in a complex and chaotic flow pattern. Turbulence in jets is responsible for increased mixing with the surrounding fluid, which is crucial in many engineering applications, such as combustion and cooling \citealp{Crow_Champagne_1971,Davies_Fisher_Barratt_1963}.
	
Despite the chaotic nature of turbulence, jets exhibit coherent turbulent structures. These structures are organized flow patterns that persist for a certain period of time and play a significant role in jet dynamics. One of the most important turbulent structures in jets is the vortex ring. 
    
The process of formation and dispersion of vortex rings is one of the main sources of turbulence in jets.\cite{Crow_Champagne_1971}
	
	In axisymmetric jets, entrainment occurs when the jet's radial spread and shear layers between the jet and ambient fluid initiate turbulence, drawing in ambient fluid. Critical to this turbulence are small, high-enstrophy structures that persist beyond the flow's characteristic timescales and primarily drive energy dissipation\cite{JIMeNEZ_WRAY_1998}.
	
	Enhanced computational power has allowed detailed examination of these structures, which generally align the vorticity vector with the main strain direction and have a typical radius five times the Kolmogorov length scale. The Kelvin-Helmholtz (KH) mechanism explains the evolution of coherent structures in shear layers and jets, representing a spatial instability that grows as it moves downstream in jets. Initially, the KH wave grows but later decays due to flow spreading, helping engineers predict and control flow in many applications.\cite{JIMeNEZ_WRAY_1998,Pickering_2020}
	
	Streamwise vortices in jets generate streaky structures through the lift-up mechanism, redistributing kinetic energy and promoting mixing. These streaks are elongated regions of alternating high- and low-momentum fluid aligned with the main flow. They extend downstream and interact with smaller-scale eddies, influencing jet turbulence. Spectral analysis shows streaky structures are prevalent in the flow's low-frequency range, highlighting their large scale and slow evolution relative to smaller turbulence, making them essential to understanding jet mixing and spreading.\cite{Nogueira19}
	
	Despite extensive research into the contributions of various  structures to the dynamics of turbulent flows, their precise roles are not yet fully understood. The development of new data-driven techniques and methodologies makes it possible to extend current knowledge, identifying potential relationships within the data that, in fact, enable a better understanding of the mechanisms underlying turbulence.
	
	\citet{Lellep2022} used the SHapley Additive explanations (SHAP) algorithm to identify the flow features relevant to predicting relaminarization, in a wall bounded flow, by analyzing the output of a machine learning (ML) model trained for this purpose. SHAP is an algorithm derived from the game theory and it quantifies the contribution of each flow feature to the model's final prediction. 

    In their work, the authors \cite{Lellep2022} showed that SHAP is an effective tool for identifying flow features influencing relaminarization and for understanding the underlying dynamic mechanisms in wall-bounded turbulence.
	
	In \citet{Cremades2024}, the authors showed that they could predict velocity fields over time using data from turbulent channel flow simulations leveraging a U-net. They then employed the SHAP algorithm, to evaluate the significance of various flow structures in making these predictions. Contrary to previous assumptions, they find that the most important structures for prediction are not necessarily those with the highest contribution to Reynolds shear stress.
	
	In the context of convolutional neural networks (CNN), explainability and interpretability Gradient-weighted Class Activation Mapping (Grad-CAM) \citep{SelvarajuDVCPB16} has emerged as a widely used technique to visualize the regions of an image that most influence the model's predictions. This method leverages the gradients flowing through the network to generate a weighted activation map, highlighting the areas most relevant to classification or regression tasks. Due to its simplicity and ability to provide intuitive visual explanations, Grad-CAM has become an essential tool for enhancing the interpretability of neural networks across various applications \citep{Panwar2020,Majid2022,Jahmunah2022}.
		
The aim of this work is to apply different XAI approaches to flow fields obtained from PIV measurements \cite{Roy2021} for an axisymmetric turbulent jet, comparing methods already used in the literature (SHAP and Gradient-SHAP) for other types of flow fields \cite{Cremades2024,Lellep2022} with GRAD-CAM, which has not yet been used in this context. This comparison of techniques aims to identify which regions of the flow field are relevant for the development of an axisymmetric turbulent jet.

\section{Methodology}\label{met_sec}
    
Convolutional Neural Networks (CNNs) have been widely employed for prediction tasks across various scientific and engineering domains, demonstrating their ability to extract meaningful patterns from complex data \cite{lecun2015deep,goodfellow2016deep}. In the context of fluid dynamics, CNNs offer a powerful tool for forecasting velocity fields, learning from past states to predict future flow evolution \cite{ling2016machine}.  

In this work, a CNN is used for a prediction task: let \( U \) represent the velocity fields on a plane at a given time \( t_i \), where a neural network is trained to predict the velocity field at the subsequent time \( t_{i+1} \). The network used is a U-Net\cite{Lecun2015}, a type of architecture that effectively extracts features at multiple spatial scales by employing feature maps of varying dimensions \cite{ronneberger2015unet}.  

During training, the neural network minimizes the loss function. If the loss function is sufficiently low at the end of training, the network is capable of accurately predicting the velocity field at the next instant, given the input at the previous instant. This suggests that the network, upon reaching a sufficiently low loss function value, has learned some of the underlying physics of the problem, making it interpretable as a model of the phenomenon itself \cite{karniadakis2021physics}.


\subsection{SHAP}\label{par:SHAP}
	
Following \cite{Lellep2022,Cremades2024} the SHAP algorithm was used to interpret and analyze the importance of variables within the adopted predictive model. SHAP, based on Shapley values from cooperative game theory, assigns a unique contribution to each feature in the model for the prediction, enabling an accurate estimate of its influence.
If $g$ is a black box function and $z$ is an input for $g$, with $z \in \mathbb{R}^M$ where $M$ as number of features, the Shapley values are defined as $\Phi_m$ \cite{Lundberg} following this eq.:
		\begin{equation}
		g(z) = \phi_0+\sum_{m=1}^{M}\phi_m
	\end{equation}
	
When dealing with interpretability tasks, it is essential to introduce the concept of a coalition: a coalition represents a subset of features considered in constructing the importance of each variable with respect to the model’s prediction. Coalitions are formed by including or excluding different features to simulate how the presence or absence of each one influences the outcome. To calculate the Shapley value, we consider all possible features' permutations. A permutation identifies a coalition, and it determines each feature's marginal contribution to every possible coalition they can join. The Shapley value for each feature is then the average of these marginal contributions over all permutations.

	
	
The Shapley values can be computed through equation \ref{eq:shap_1}:
	
	\begin{equation}\label{eq:shap_1}
		\phi(i, v)=\sum_{S \subseteq N \backslash\{i\}} \frac{|S|!(|N|-|S|-1)!}{|N|!}[v(S \cup\{i\})-v(S)]
	\end{equation}
	
\noindent where \( \phi(i,v) \) is the SHAP value due to the \(i\)-th feature, \( N \) is the set of coalition participants, \( S \) is a subset of features excluding \( i \), \( v(S \cup \{i\}) - v(S) \) is the marginal contribution term for the \( i \)-th feature and \( \frac{|S|!(|N| - |S| - 1)!}{|N|!} \) is the probability that the \( i \)-th feature will be part of the coalition \( S \)\cite{Lundberg}.
	
The exact calculation of Shapley values is not immediate; therefore, the typical approach is to approximate them with the kernel-SHAP method \cite{Lundberg,covert21}, which integrates Shapley values with the Local Interpretable Model-agnostic Explanations (LIME) method \cite{riberio16}. This can be expressed as follow,
	
\begin{equation}
		\mathcal{L}\left(f, g, \pi_x\right)=\sum_{d=0}\left[f\left(h_x\left(q^{\prime}\right)\right)-g\left(q^{\prime}\right)\right]^2 \pi_x\left(q^{\prime}\right)
	\end{equation}
	
	\begin{equation}
		\pi_x\left(q^{\prime}\right)=\frac{|Q|-1}{\binom{|Q|}{q^{\prime} \mid}\left|q^{\prime}\right|\left(|Q|-\left|q^{\prime}\right|\right)}
	\end{equation}
	
\noindent where $|q'|$ is the number of structures, $h_x$ represents a mask function that tranforms the binary space of $q'$ into the space of the input of the model. The LIME equation has been solved by means of a linear regression, with the resulting error being $(f-g)^2 \mathcal{O}(10^{-7})$ \cite{riberio16}.
	
The local accuracy property of Shapley values guarantees that the total of the Shapley values for a given sample \( z \) equals the difference between the model's output for that sample, \( g(z) \), and the average prediction of the model, \( \mathbb{E}[g(\tilde{z})] \). Thus, the total of all Shapley values is equivalent to the difference between the model output and the average prediction of the model.
	
While the application of SHAP to regression or classification problems is straightforward \cite{das2020}, in cases like the one analyzed in this work, flow prediction, it becomes more complex because the input and/or output lack predefined features. For example, in \citet{Cremades2024}, the authors defined the structures present in the field leveraging the approach proposed by \citet{Lozano2012}, and using each of these structures as features for calculating their importance via SHAP.
	
In this work, we focus on an axisymmetric jet flow. We have chosen to use the vorticity normal to the symmetry plane, \( \omega_z \), as a discriminant to determine input features, implementing a thresholding method.
	
We used the jet exit diameter $D$ and $V_{exit}$ to normaliz \( \omega_z \):
	\begin{equation}
		\omega_z^* = {\omega_z \cdot D \over V_{exit}}
	\end{equation}
	
Specifically, a threshold \( \omega_{\text{thr}} \) is set and applied to the flow field, yielding two distinct masks: for \( \omega_z^* > \omega_{\text{thr}}^* \) and \( \omega_z^* < -\omega_{\text{thr}}^* \), respectively For each mask, the connected regions are labeled, and these regions serve as input features, as sketched in Figure \ref{fig:schema}.
	
\subsection{Gradient-SHAP}\label{par:gradientshap}
	
Gradient-SHAP is a method for attributing feature importances to the predictions of a machine learning model \cite{sunda_17,Erion2021}. It combines the ideas of SHAP with the gradients of a model outputs with respect to the input features. This method aims to explain the outputs of complex models, such as deep neural networks, by providing robust explanations even in highly nonlinear settings.
	
The choice of \( \mathbf{x}' \) plays a critical role in defining the reference from which contributions are measured, aligning with the broader SHAP methodology. Given a model \( f \) and an input vector \( \mathbf{x} \), Gradient SHAP estimates the contribution of each feature \( x_i \) to the model output \( f(\mathbf{x}) \) by considering a baseline input \( \mathbf{x}' \); in the present case we consider the mean flow velocity. The method interpolates between \( \mathbf{x} \) and \( \mathbf{x}' \) by generating random samples \( \mathbf{z} \) along a linear path, defined as:
	
	\begin{equation}
		\mathbf{z}_\alpha = \alpha \mathbf{x} + (1 - \alpha) \mathbf{x}', \quad \alpha \in [0, 1]
	\end{equation}
	
The contributions are then estimated by sampling multiple values of \( \alpha \) and computing the gradients of \( f \) with respect to \( \mathbf{z}_\alpha \). The SHAP value for the \( i \)-th feature is calculated as:
	
	\begin{equation}
		\phi_i = \mathbb{E}_{\alpha \sim \text{Uniform}[0,1]} \left[ (\mathbf{x} - \mathbf{x}')_i \cdot \nabla_{\mathbf{z}_\alpha} f(\mathbf{z}_\alpha) \right]
	\end{equation}
	
\noindent where \( \nabla_{\mathbf{z}_\alpha} f(\mathbf{z}_\alpha) \) denotes the gradient of the model output with respect to \( \mathbf{z}_\alpha \).

\subsection{GradCAM}\label{par:GradCAM}
	
The Grad-CAM algorithm, introduced by \citet{SelvarajuDVCPB16}, represents a significant advancement over previous approaches regarding versatility and accuracy in interpretability tasks, but it is applicable exclusively to neural networks. The method can be introduced at a high level in a very simple way: the process begins by taking a input image and constructing a model that is truncated at the layer of the neural network for which we want to create a Grad-CAM heat-map following the algorithm shown below. It is important to point out that the layer against which to truncate the network, represents a parameter of the algorithm. For prediction purposes, fully connected layers are added. Subsequently, the input traverses the model, capturing the output and the loss. Then, the gradient of the output of the model layer with respect to the model loss is calculated. Next, we identify the portions of the gradient contributing to the prediction, refining, resizing, and adjusting their scale to overlay the heatmap onto the original image.
	
The main idea behind Grad-Cam is that the convolutional layers naturally preserve spatial information that is lost in fully-connected layers \cite{SelvarajuDVCPB16,Vinogradova_2020}. \citet{Vinogradova_2020} used Grad-CAM for a CNN trained for Semantic Segmentation and they suggest choosing the last convolutional layers to strike the best balance between high-level semantics and detailed spatial information. Neurons in these layers seek out semantic class-specific information in the image. Grad-CAM utilizes the gradient information in input to the last convolutional layer of the CNN to assign importance values to each neuron for a particular decision of interest.
	
Following the original implementation \cite{SelvarajuDVCPB16}, we consider a CNN with an RGB image as input (\(\text{Img} \in \mathbb{R}^{x \times y \times 3}\)), where \(x\) and \(y\) represent the width and height of the image, respectively. The output is a tensor \(\in \mathbb{R}^{x \times y \times c}\), where \(c\) represents the number of classes. 

The first operation involves computing the gradient (Eq. \ref{eq:eq_alpha}) for each class with respect to the feature map activations \(A^k\) of a convolutional layer, where \(K\) is the number of feature maps (kernels) in the target layer. These gradients are then averaged across the spatial dimensions of the image to obtain an importance weight \(\alpha_k^c\). The coefficient \(\alpha_k^c\) represents a partial linearization of the deep network downstream of \(A^k\) and captures the importance of the feature map \(k\) for a target class \(c\).

\begin{equation} \label{eq:eq_alpha}
    \alpha_k^c = \overbrace{\frac{1}{Z} \sum_m \sum_n}^{\text{Global Average Pooling}} 
    \underbrace{\frac{\partial y^c}{\partial A_{mn}^k}}_{\text{Gradients via backprop}}
\end{equation}

where \(Z = x \times y\) is a normalization factor corresponding to the total number of spatial locations in the feature map.

The last step is the weighted combination of activation maps followed by a ReLU to obtain eq. \ref{eq:relu_eq_alpha}.
\citet{SelvarajuDVCPB16} introduce the ReLU in eq. \ref{eq:relu_eq_alpha} because they are only interested in the features that have a positive influence on the class of interest, i.e. the pixels whose intensity should be increased to increase $y^c$.
	
	\begin{equation}\label{eq:relu_eq_alpha}
		L_{\text {Grad-CAM }}^c=\operatorname{Re} L U \underbrace{\left(\sum_k \alpha_k^c A^k\right)}_{\text {linear combination }}
	\end{equation}
	
This methodology can be applied to any differentiable activation function, as reported in \cite{SelvarajuDVCPB16}.
	
In \citet {Vinogradova_2020}, the authors adapted Grad-CAM to an image segmentation problem, by substituting $y^c$ in eq \ref{eq:eq_alpha} with proposed $\sum_{(m, n) \in {M}} y_{m n}^c$, where $M$ is a set of pixel indices of interest in the output mask. $M$ can denote just a single pixel, or pixels of an object instance, or simply all pixels of the image. 
	
	What is discussed in this paper starts from the approach proposed by \citet{Vinogradova_2020} in order to extend Grad-CAM to a flow-prediction problem, replacing flow quantities, i.e. $y = u$.
	
	\section{Results}
	
\begin{figure}
		 		\centering
		
		{\includegraphics[width=0.75\columnwidth]{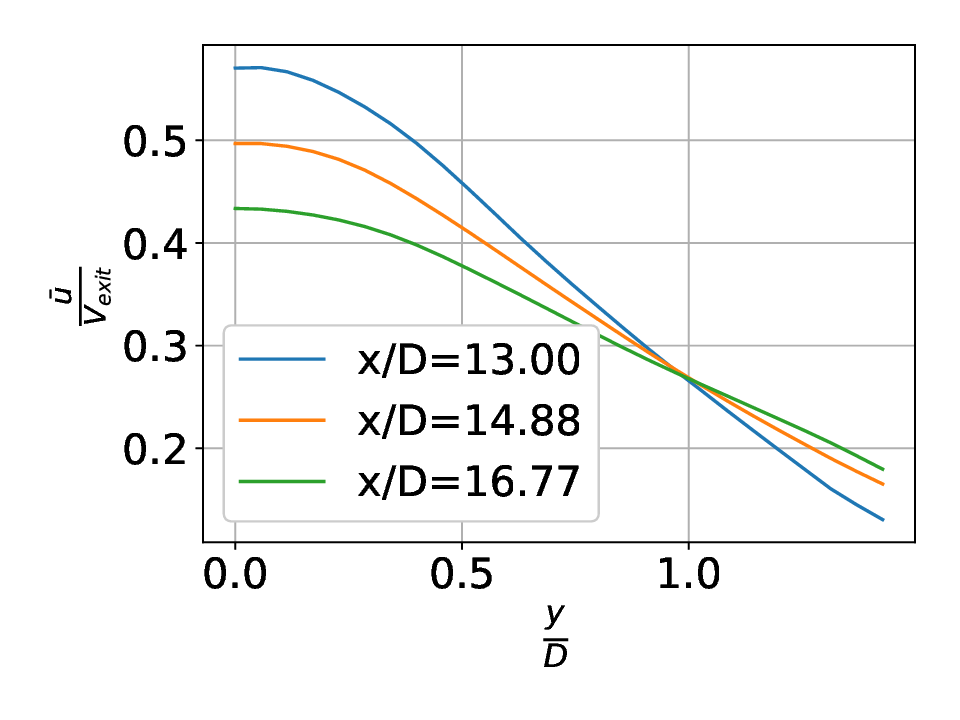}}
		
		\caption{Time-averaged streamwise velocity component normalised with the jet's exit velocity ($V_{exit}$) from dataset\cite{Roy2021} plotted as a function of the spanwise coordinate.}
		\label{fig:mean_field}
		
	\end{figure}

The dataset \cite{Roy2021} consists of 8,451 instantaneous velocity snapshots, obtained via PIV of an axisymmetric jet. The analyzed jet flow is characterized by a Reynolds number (Re), based on the mass flow rate and the jet diameter, of \( Re = 21{,}000 \), with an exit velocity \( V_{\text{exit}} = 66 \). The spatial mesh grid comprises 67x51 points, spaced in both directions at \( \Delta z = \Delta x = 0.268 \, \text{mm} \). The time resolution is \( \Delta t = 0.01 \, \text{ms} \). The streamwise velocity profiles along the spanwise direction for $x/D = 13, 14.88$, and $16.77$ are shown in Figure \ref{fig:mean_field}.

To extract meaningful features from this dataset according methods described in \ref{par:SHAP},\ref{par:gradientshap}, \ref{par:GradCAM}, we employ a deep learning approach based on a U-Net architecture.

\begin{figure}
		\centering
		\includegraphics[width=\columnwidth]{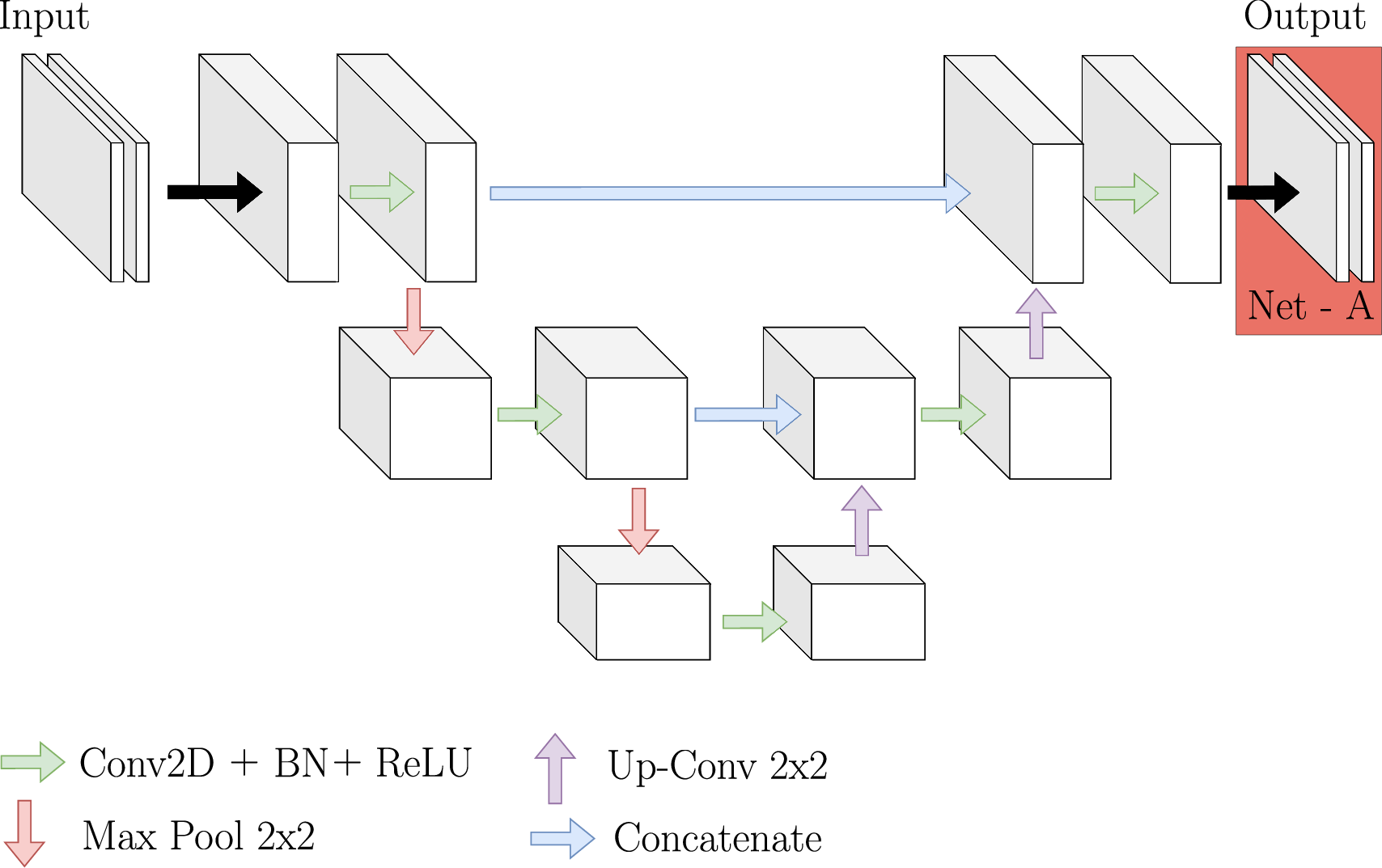}
		\caption{Scheme on the implemented neural network.}
		\label{fig:unet}
\end{figure}

The implemented U-Net architecture consists of three main components, as sketched in Figure \ref{fig:unet}: the contracting path (encoder), the bottleneck, and the expansive path (decoder). This architecture is designed to learn hierarchical representations of input data by combining local and global information through skip connections, as described in \citet{ronneberger2015unet}.

The encoder comprises a series of convolutional blocks, each performing a two-dimensional convolution followed by batch normalization to stabilize and accelerate the learning process. These operations are followed by the ReLU activation function, which introduces non-linearity and enhances the model's expressiveness. At the end of each block, a MaxPooling2D operation with a \(2 \times 2\) kernel is applied, reducing the spatial dimensions of the feature maps by a factor of 2. This enables the network to extract increasingly non-linear features as it progresses through the layers.

The bottleneck represents the deepest part of the U-Net and serves as a latent space that captures the most compact and meaningful representations of the input data. This stage is crucial for learning high-level non-linear features.

The decoder is designed to progressively reconstruct the spatial dimensions of the image. It uses transposed convolutions with a \(2 \times 2\) kernel to double the spatial dimensions of the feature maps. In each block of the decoder, the resulting feature maps are concatenated with those from the corresponding block in the contracting path via skip connections, allowing the network to combine high-level semantic details with low-level, high-resolution information. After concatenation, convolutions, batch normalization, and ReLU activations are applied to further refine the reconstructed features.

The final output of the network is obtained by applying a convolution that adjusts the number of channels to match the desired output, ensuring that the resulting tensor has the same spatial dimensions as the input. This modular and flexible structure makes the U-Net particularly effective for image segmentation and reconstruction tasks, preserving important details by integrating local and global information.

The skip connections, in particular, play a critical role in enhancing the network's performance by enabling direct information flow across layers, as highlighted in previous studies \cite{ronneberger2015unet, drozdzal2016importance}.

The NN was trained randomly splitting the dataset into a training (60\%) and a validation set (40\%). 

For the training process, a RMSprop gradient-descent algorithm was used \cite{Zou2018ASC}, with a learning rate of $10^{-4}$ and a momentum of $0.9$. We used the early stopping in order to prevent overfitting. At the end of the training the rms of the error was around $10^{-4}$.

\begin{figure}
		\centering
		\includegraphics[width=.5\columnwidth]{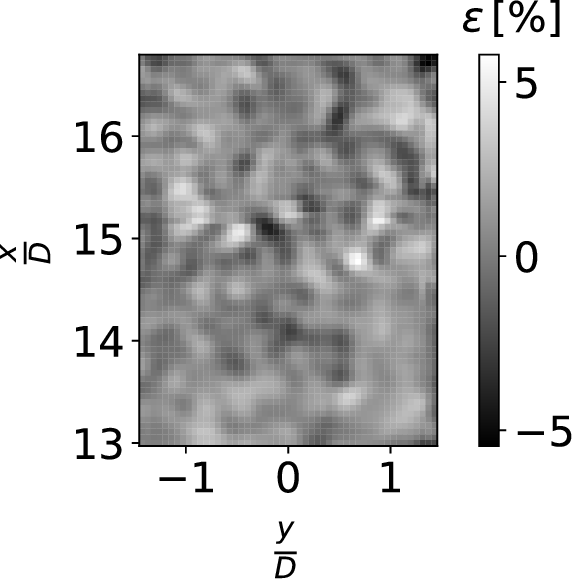}
            \caption{Percentage error on the velocity prediction}
		\label{fig:error}
\end{figure}

Figure \ref{fig:error} shows the percentage error ($\epsilon$) of a velocity field predicted by the trained NN. The error is calculated as the local difference with respect to the ground truth of the predicted velocity field, normalised with respect to the ground truth. The error is quite limited  to within $\pm 4 \%$and as such is of minimal consequence for the intended application of the neural network.
\begin{figure*}
		\centering
		\includegraphics[width=1.5\columnwidth]{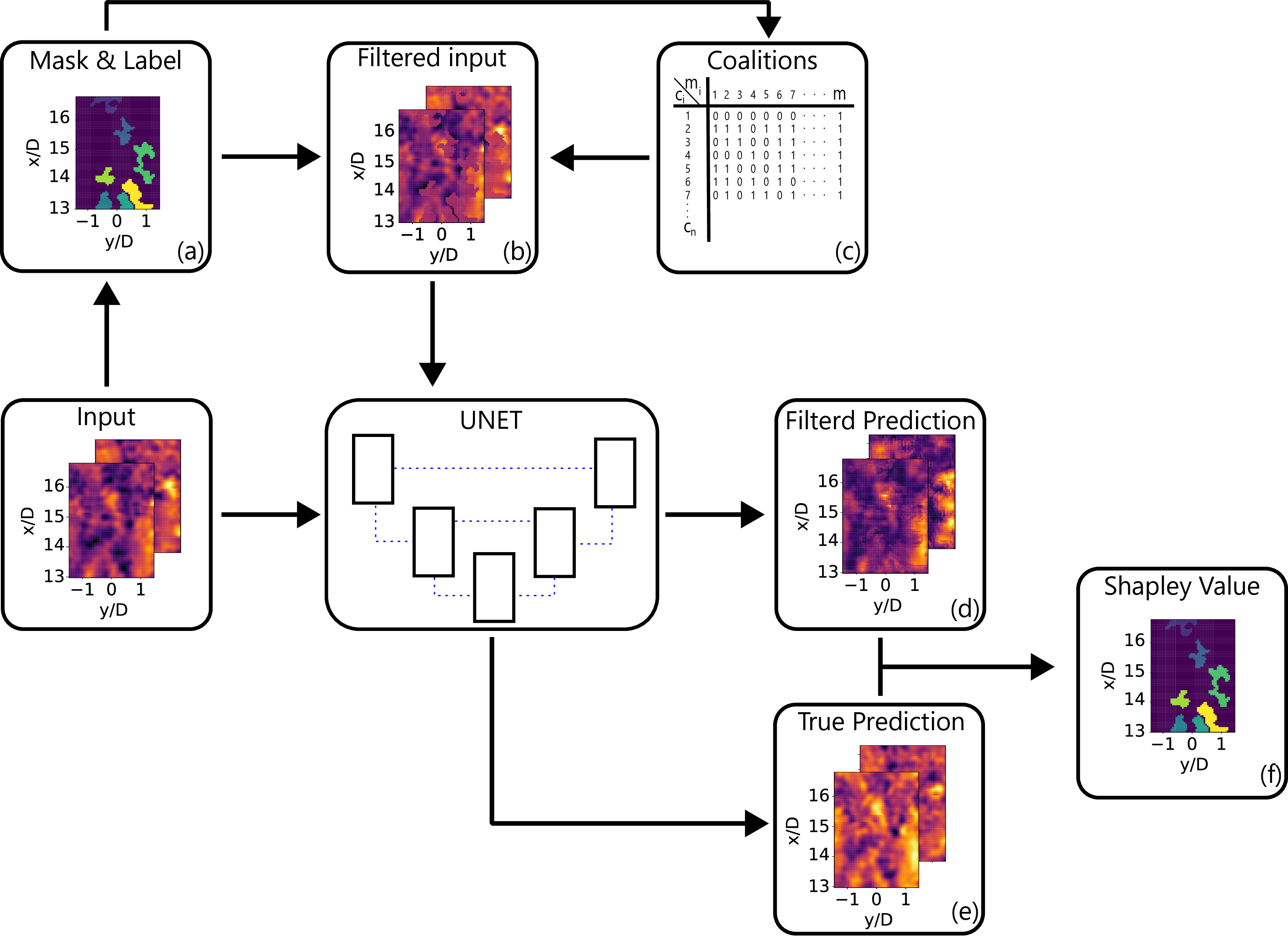}
		\caption{Schematic representation of the framework used to calculate the Shapley values. For each validation snapshot, a filter is applied based on vorticity $\omega_z^*$ to identify the structures in the snapshot (a). For each snapshot, coalitions are then constructed (c), which determine which structures to remove from the input snapshot (b). The filtered image is then provided as input to the neural network, and the obtained output (d) is compared with (e), in order to obtain the Shapley values.}
		\label{fig:schema}
\end{figure*}

The prediction task performed by the U-net is implemented in fed framework used to extract the features’ importance, as schematically described in Figure \ref{fig:schema}. As reported in the section \ref{met_sec}, a thresholding on the value of $\omega_z$ is applied to each snapshot to identify the connected regions within each frame. The resulting identified regions are then labelled, as indicated in Figure \ref{fig:schema}a. Coalitions are formed by including or excluding different features to simulate how the presence or absence of each one influences the predictive outcome (see Figure \ref{fig:schema}c).
	
For each coalition element, the corresponding filtered input is calculated. This input is identical to the real input, except that elements not in the coalition are set to zero. The filtered input is then provided to the neural network, which performs the prediction. The same process is repeated with the unfiltered input. By comparing the two outputs, it is possible to measure how important the coalition is for the prediction.

\begin{figure}
		\centering
		\includegraphics[width=.5\columnwidth]{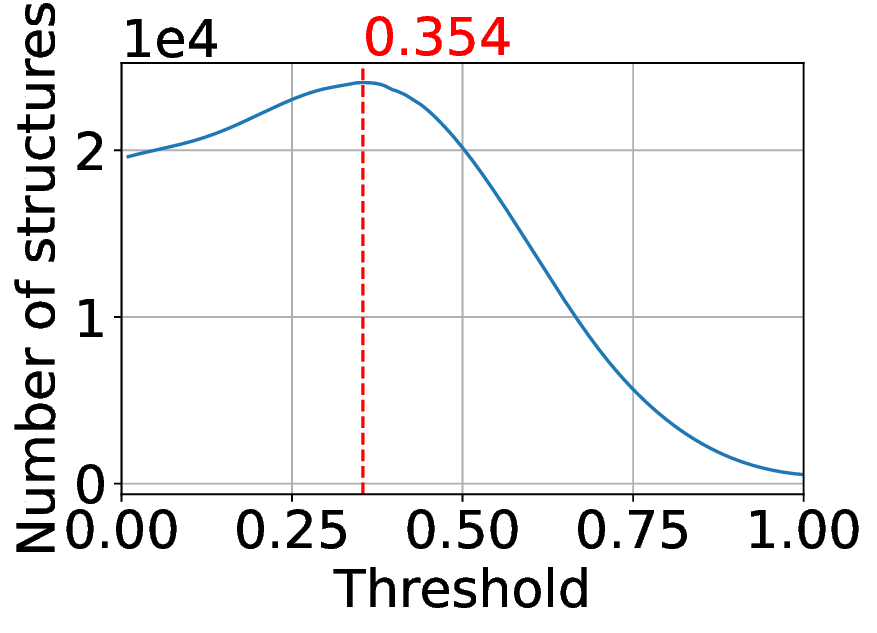}
		\caption{Results of the percolation analysis. The red dotted line highlights the selected value of $\omega_z^*$ used for the filter.}
		\label{fig:percol}
\end{figure}

The outcome can be sensitive with respect to the choice of \( \omega_{\text{thr}}^* \), as this obviously influences the identification. Therefore, a percolation analysis was performed to select the optimal value that maximizes the number of identified structures, as shown in Figure \ref{fig:percol}. The resulting value of $\omega_{thr}^*$ is 0.354.
	 
\begin{figure}		\centering
		\subfloat[][]
		{\includegraphics[width=.5\columnwidth]{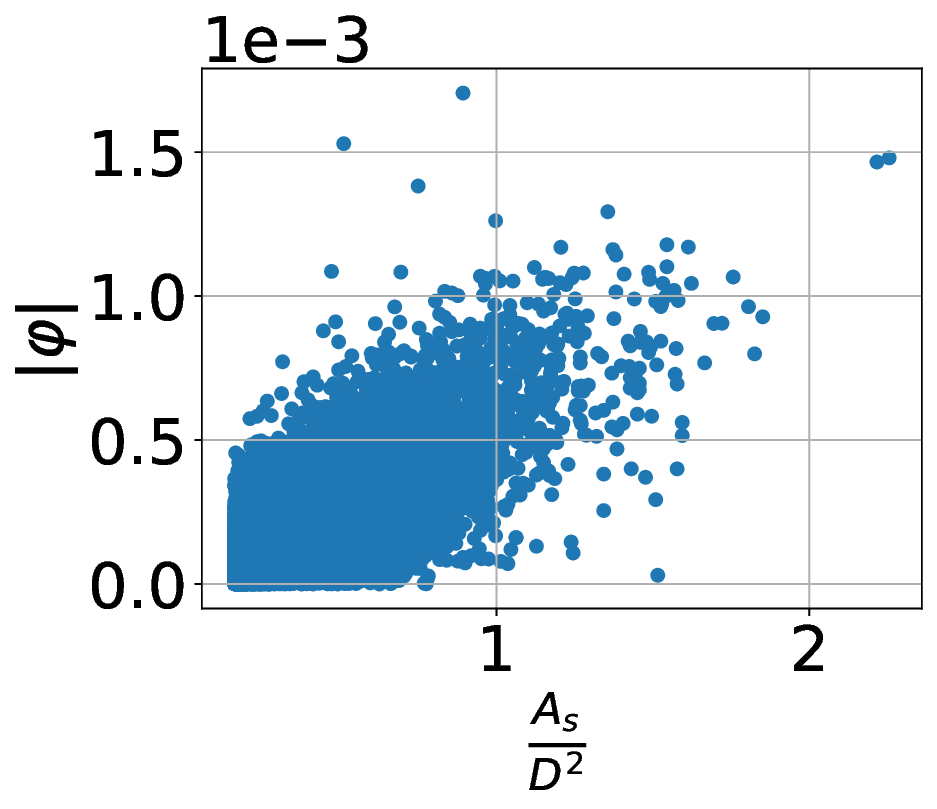}}
		\subfloat[][]
		{\includegraphics[width=.5\columnwidth]{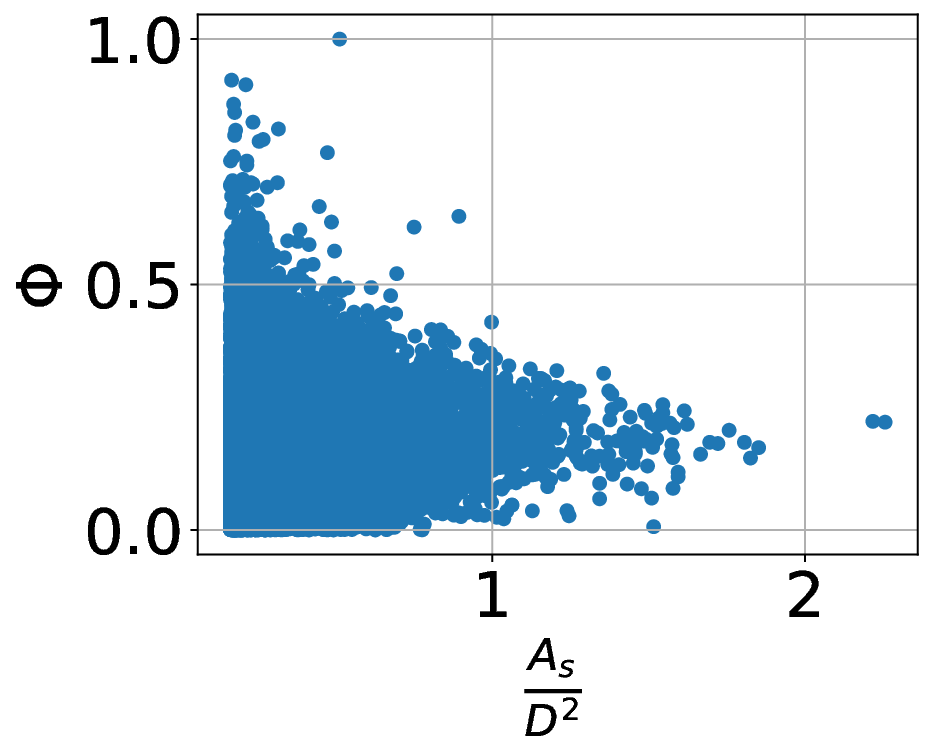}}\\
		
\caption{Scatter plot of the Shapley values (a) and normalized Shapley values (v) versus the corresponding area ($A_s$), normalized by the jet exit diameter ($D$), for each identified structure.}
\label{fig:shap_area}
\end{figure}

Figure \ref{fig:shap_area}a shows the values of \( \phi \) as a function of the area of the structure. As the area increases, \( \phi \) increases accordingly. This is an obvious consequence of the fact that the larger the area of a turbulent structure, the greater content will be removed from the snapshot, or more specifically the greater the amount of information that will be removed from the input data. Consequently, the network will produce a less accurate result, and the Shapley value will increase. \citet{Cremades2024} already showed that this is an obvious result, which does not add any significant information to the interpretation of the flow.
	
To avoid this misleading result, a normalization is introduced according to the area occupied by the structure ($A_s$), specifically defined as:
	
	\begin{equation}
		\Phi = \frac{|(\phi / A_s)|}{\max(|\phi / A_s|)}
	\end{equation}
	
Figure \ref{fig:shap_area}(b), shows that the most important structures are not necessarily those with high values of \( A_s \). To better understand what the identified structures represent in relation to their importance, let us consider evaluating an integral value of vorticity on each structure. 
An integral value of the vorticity calculated over the identified structures, with area As, allows us to draw some conclusions on the feature importance. The average is calculated as: 

\begin{equation}\label{eq:mea_int}
		<f> = {1 \over A_s}{\int\limits_{A_s} f(x,y)d\sigma} 
\end{equation}

Where f indicates the generic quantity that is being analysed.

\begin{figure}
		\centering
		\subfloat[][]
		{\includegraphics[width=.525\columnwidth]{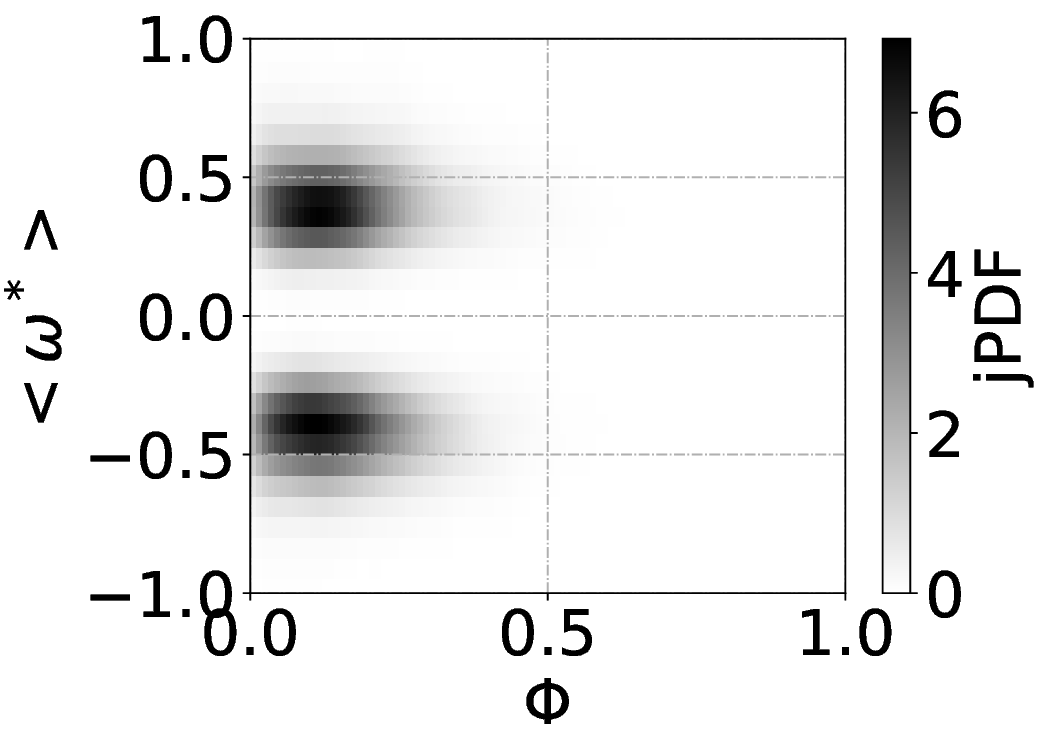}}
		\subfloat[][]
		{\includegraphics[width=.475\columnwidth]{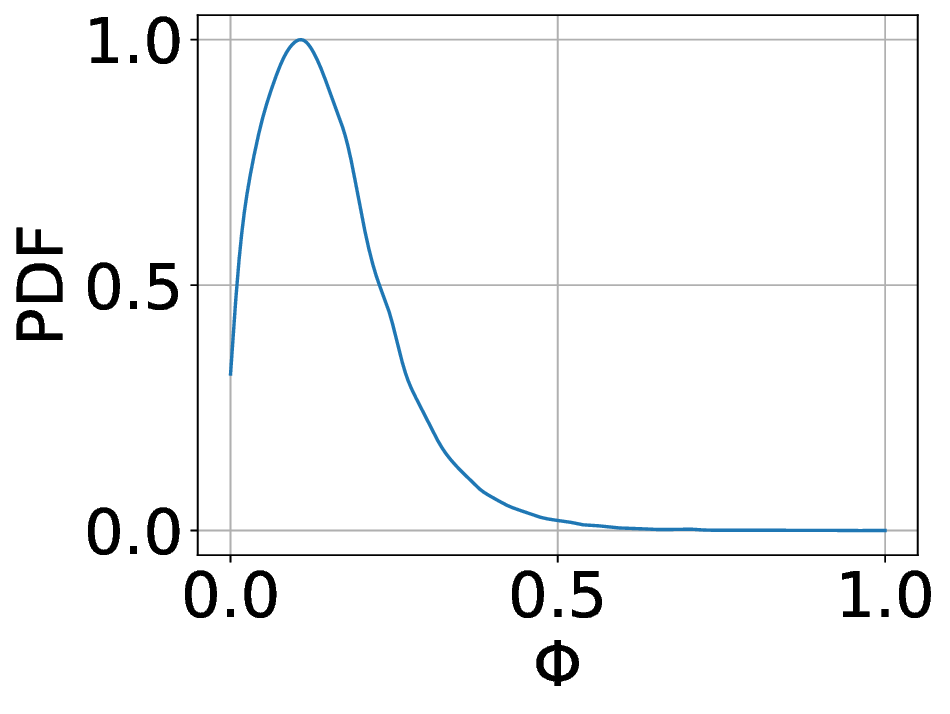}}\\
		
		\caption{Joint PDF of normalized Shapley values vs average integral $\omega_z$ (a). PDF of the average integral of normalized Shapley values (b).}
		\label{fig:cluster}
\end{figure}

Figure \ref{fig:cluster}a shows the joint probability density function (jPDF)  between \( \Phi \) and \( \langle \omega \rangle \), which highlights two main clusters, based on the sign of the \( \langle \omega_z \rangle \).
		
It is evident that, regardless of the sign of \( \langle \omega_z \rangle \), higher values of \( \Phi \) occur for \( 0.25 < | \langle \omega_z \rangle | < 0.5 \). This suggests, contrary to initial assumptions, that the most relevant regions are not those with high vorticity.

Additionally, it is clear that most of the identified structures show values of \( 0.1 < \Phi < 0.2 \). 
It is important to note that even though \( \Phi \) is normalized to reach unity, as shown in Figure \ref{fig:cluster}b, the probability of \( \Phi > 0.5 \) is negligible. Therefore, it can be concluded that statistically, the identified structures have \( 0 < | \langle \omega_z \rangle | < 0.75 \). It is highly probable to find moderately important structures (\( \Phi \sim 0.15 \)) with \( 0.25 < | \langle \omega_z \rangle | < 0.5 \).

\begin{figure}
 		\centering
 		\subfloat[][]
 		{\includegraphics[width=.5\columnwidth]{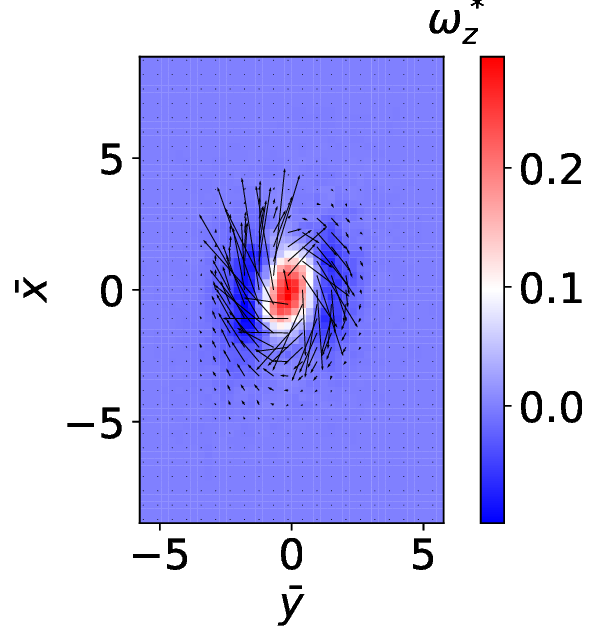}}
 		\subfloat[][]
 		{\includegraphics[width=.5\columnwidth]{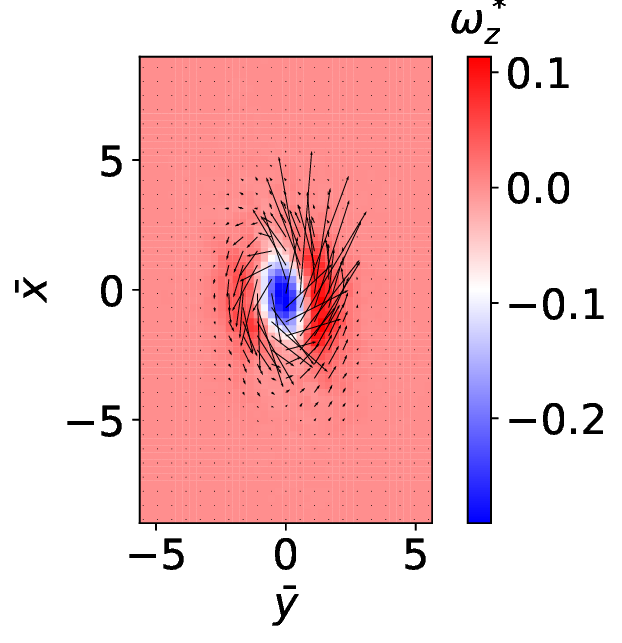}}\\
 		
 		\caption{Conditional average structures with the value of $\omega_z^*>0$ (a) and $\omega_z^*<0$ (b), superimposed there soon the velocity vectors u and v.}
 		\label{fig:struct_cond}
\end{figure}

In figures \ref{fig:struct_cond}a,\ref{fig:struct_cond}b are shown the conditionally averaged structure corresponding to the two clusters characterised by values of positive and negative vorticity, respectively.
Specifically, for each region identified using the \( \omega_{\text{thr}}^* \) filter, the bounding box containing it, its center, and all structures were identified. In accordance with their cluster, they were re-centered to share a common center and then averaged. The two clusters have a comparable number of structures, which is to be expected, given the axisymmetric flow field. This property is reflected in the conditionally averaged structures, which show a similar topology, as seen in Figures \ref{fig:struct_cond}a and \ref{fig:struct_cond}b, with an inverted rotation direction. The conditionally averaged structure shows an extent of approximately ${{x}\over{D}} \sim 5$ in the two dimensions considered. 
	
\begin{figure}
 		\centering
 		\includegraphics[width=.5\columnwidth]{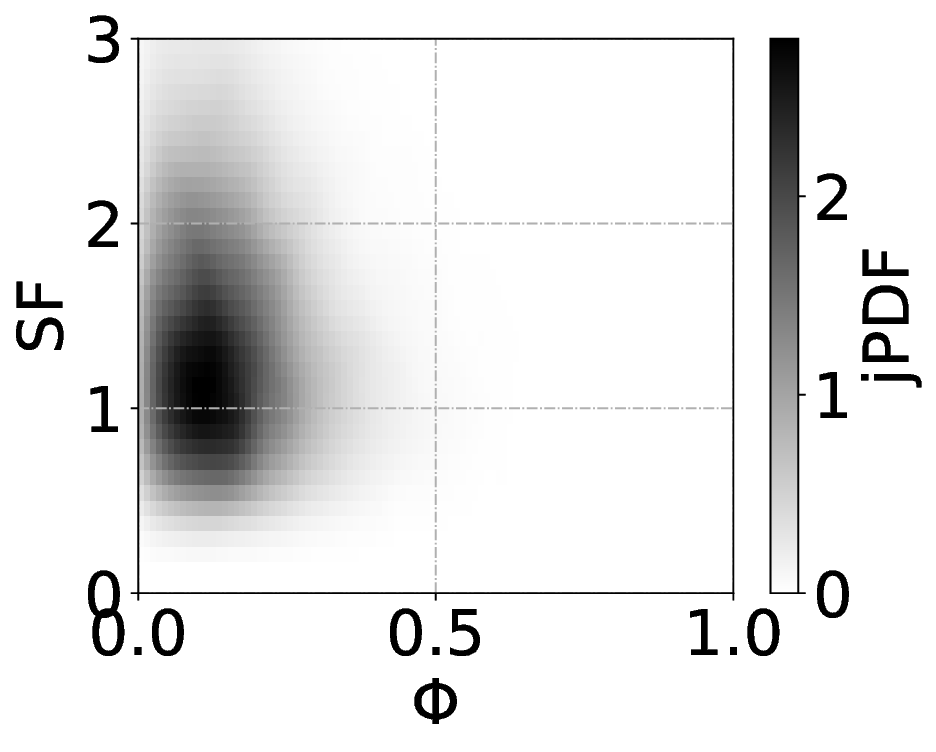}
 		\caption{Joint PDF of Shape Factor and Shapley values.}
 		\label{fig:jpdf_shape}
\end{figure}

The jPDF between the shape factor (SF), defined as the ratio between the extent of the bounding box containing the structure in the streamwise direction and in the spanwise direction, and $\Phi$ is shown in Figure \ref{fig:jpdf_shape}. It is evident that structures with $SF = 1$ are not only the most common but also have the highest values of $\Phi$.

Besides the connection with the topology of the flow structures, it is also of interest to understand how these structures and their importance evaluated with the Shapley values can be related to the turbulence properties of the flow. We introduce the turbulence dissipation:
	\begin{equation}
		\varepsilon = -\nu\overline{{\partial u_i' \over \partial x_j}{\partial u_i' \over \partial x_j}}
	\end{equation}

where $u' = u - \overline{u}$ is obtained through the Reynolds decomposition and subscript "$i"$ denotes the components.

\begin{figure}
	\centering
	\includegraphics[width=.8\columnwidth]{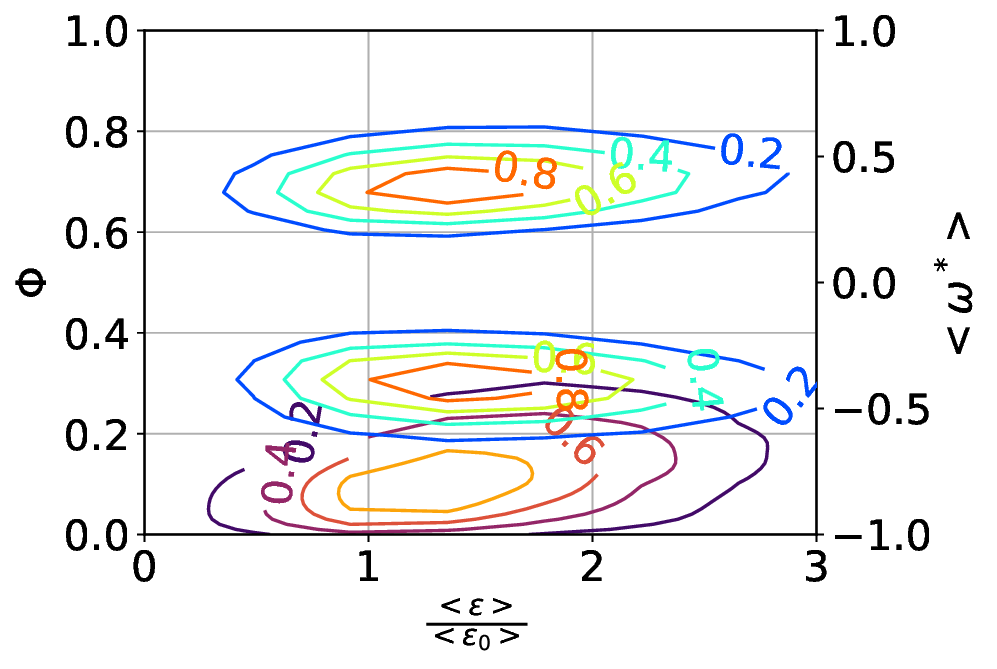}
	\caption{Joint PDF of the normalised dissipation function and the normalised
		Shapley values is illustrated in the dark lines of the contour. Joint PDF of the normalised dissipation function and the integral mean of the normalised vorticity is displayed in the light contour lines.}
    \label{fig:eps_}
\end{figure}

The average integral,obtained with eq. \ref{eq:mea_int}, of the dissipation function on the structure’s surface $<\epsilon>$ was normalized with the average integral of the dissipation function’s computed on the entire field $<\epsilon_0>$.
From Figure \ref{fig:eps_}, it can be seen that as \( \frac{\langle \epsilon \rangle}{\langle \epsilon_0 \rangle} \) increases, \( \Phi \) also increases. This suggests that the most important structures are those that contribute the most to dissipation.

Structures with high values of $\langle \varepsilon \rangle / \langle \varepsilon_0 \rangle$ show a clear correlation with an increase in $\Phi$, suggesting that dissipation is an effective measure for identifying the most relevant regions. From Figure \ref{fig:eps_} it is inferred that a concentration of dissipation near areas of moderate vorticity, consistent with studies such as those by \cite{Hussain_1986}. In fact \cite{Hussain_1986} points out that dissipation is not concentrated in regions of maximum vorticity but rather in adjacent areas where the interaction between coherent structures occurs, and where longitudinal vortices are stretched and connect to spanwise vortices. Moderate vorticity areas, in this context, refer to the saddle regions between coherent structures, where most turbulence production and dissipation takes place.
	
\begin{figure*}
		\centering
		\includegraphics[width=1.5\columnwidth]{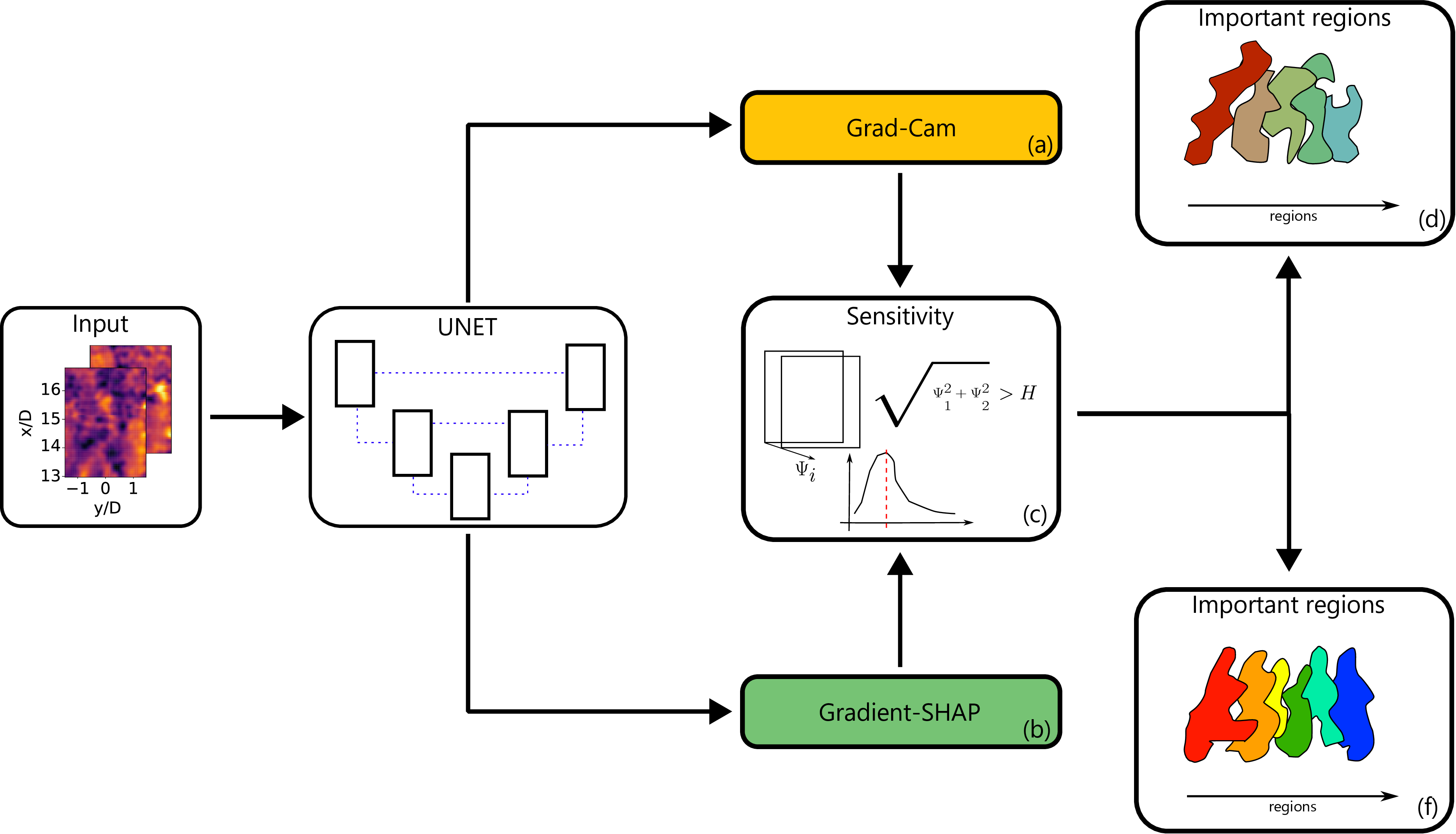}
		\caption{Schematic representation of the framework used for Grad-CAM (a) and Gradient-SHAP (b). Each of the two algorithms is applied to the trained UNET. After applying Gradient-SHAP and Grad-CAM for each validation snapshot of the network, a percolation analysis is performed on the resulting heatmaps (c) to identify the important regions within the flow field (d) and (f).}
		\label{fig:schema2}
\end{figure*}

As explained in section \ref{met_sec}(A) and (B), the advantage of using Grad-CAM and Gradient-SHAP lies in not having to predefine a metric to infer the important regions in the flow, since these approaches are agnostic. This allows for no threshold removing the assumptions and determining a posteriori which regions are significant to the flow. As represented in Figure \ref{fig:schema2}, the UNET only takes in input the raw snapshots, since no filtering is applied to them. 

The schematic framework of Grad-CAM and GRADIENT-Shap is shown in Figure \ref{fig:schema2}. These methods give in output heatmaps ($\Psi_i$) that carry the importance of each input pixel on the final prediction.  The subscript $i$ for Grad-CAM depends on the number of output channels, while for Gradient-SHAP it depends on the number of input channels. In the present case, the prediction maps the velocity components ($u$, $v$) to the same velocity components ($u$, $v$), thereof resulting in $i=2$ in both cases.

\begin{figure}
	\centering
	\subfloat[][]
	{\includegraphics[width=.5\columnwidth]{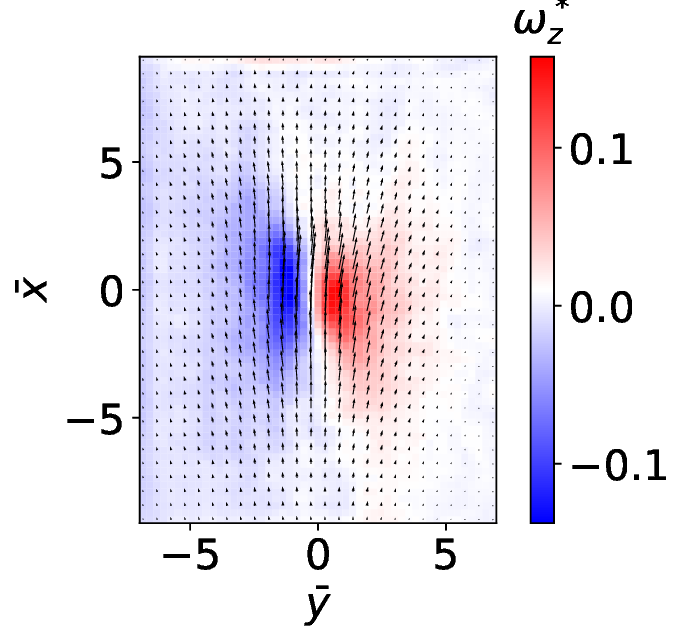}}
	\subfloat[][]
	{\includegraphics[width=.5\columnwidth]{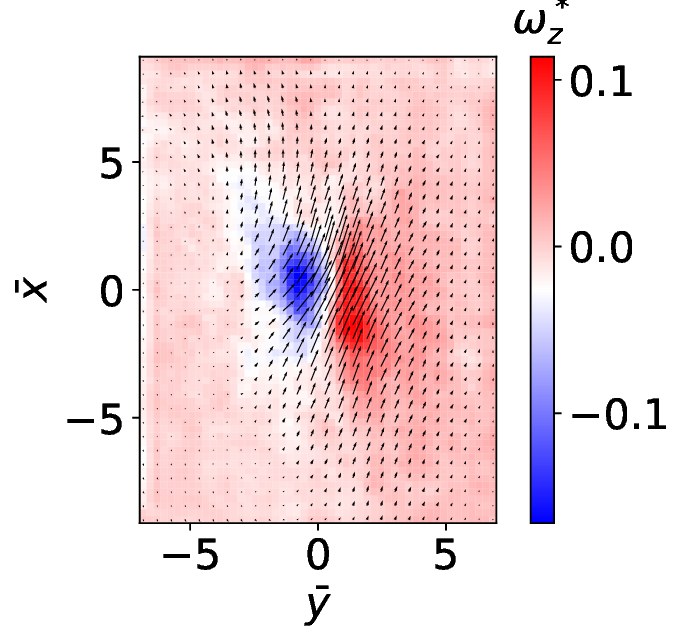}}\\
	
	\caption{Grad-CAM conditional average structures with the value of $\omega_z^*>0$ (a) and $\omega_z^*<0$ (b), superimposed there soon the velocity vectors u and v.}
	\label{fig:struct_cond_GRACAM}
\end{figure}

\begin{figure}
	\centering
	\subfloat[][]
	{\includegraphics[width=.5\columnwidth]{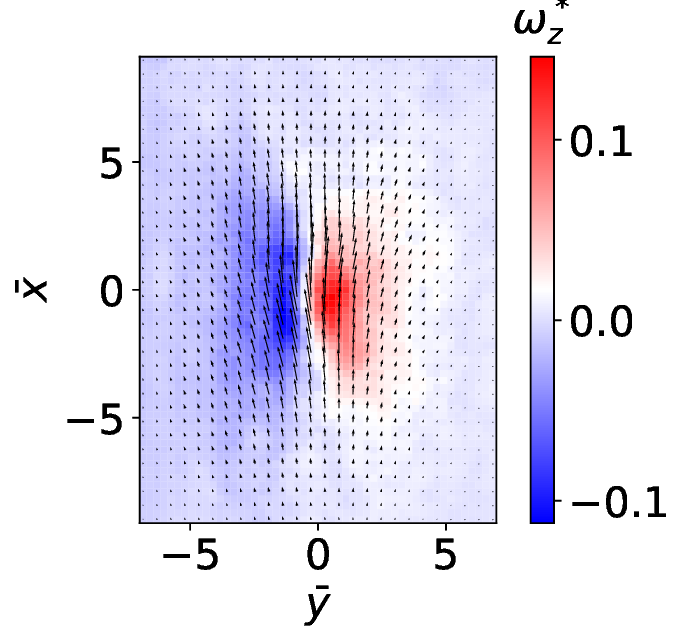}}
\subfloat[][]
{\includegraphics[width=.5\columnwidth]{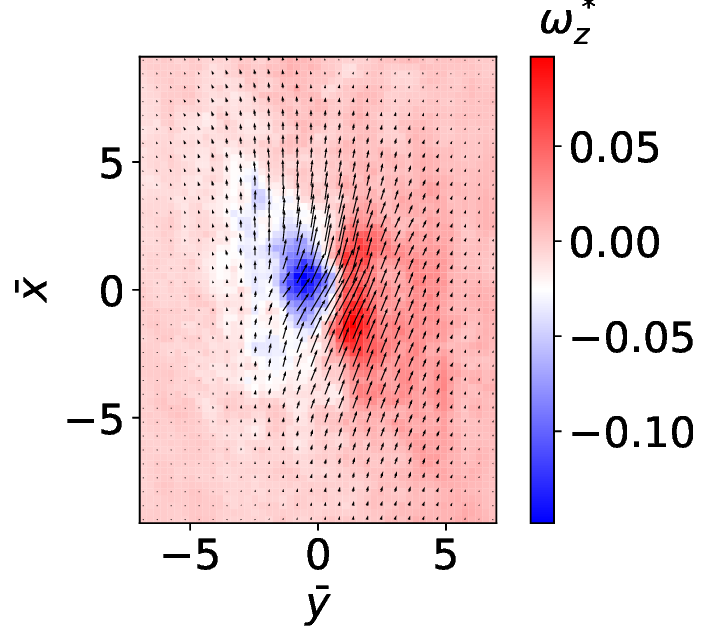}}\\

	\caption{Gradient SHAP conditional average structures with the value of $\omega_z^*>0$ (a) and $\omega_z^*<0$ (b), superimposed there soon the velocity vectors u and v.}
	\label{fig:struct_cond_GRADIENT}
\end{figure}

To extract relevant regions of the flow, a filter is applied to $\Phi_i$ to identify connected regions. The applied filter requires that the heatmap values are such that, at a pixel level, it must be verified that $\sqrt{\Psi_1^2+\Psi_2^2}>H$. The value of the threshold $H$ is chosen through percolation analysis to maximize the number of identified structures, and it is set to 0.87 for Grad-CAM and 0.3 for Gradient-SHAP. 

The regions of the flow that were identified with Gradient-SHAP and Grad-CAM according to the criterion defined above, are those that are relevant to the flow. However, there is no pre-filtering to the input data, as it was done in the case of Shapley. 
In order to compare the results obtained with the three methods, a common metric is defined based on the out-of-plane vorticity. In particular, we focus on the conditionally averaged structures similar to those identified in Figure \ref{fig:struct_cond}. 
Figures \ref{fig:struct_cond_GRACAM} and \ref{fig:struct_cond_GRADIENT} display the average conditional structures for $\omega_z>0$ (see Figures \ref{fig:struct_cond_GRACAM}a and \ref{fig:struct_cond_GRADIENT}a), and for $\omega_z<0$ (see Figures \ref{fig:struct_cond_GRACAM}b and \ref{fig:struct_cond_GRADIENT}b). An initial evident result emerges by comparing these average conditional structures with Figure \ref{fig:struct_cond}: indeed, Grad-CAM and  Gradient-SHAP yield similar results in terms of conditional structures, and they are both different from those obtained with SHAP. This is not surprising, since the results from SHAP are forced to identify those regions that are characterized by values of ($|\omega_z|>{\omega_z}_{thr}$), which inevitably leads to the identification of the vertical structures shown in figure \ref{fig:struct_cond}. This bias could hinder the generality of the results, somehow hiding the physical mechanism at play. 

The physics of turbulent jets involves energy transfer across multiple scales, from large vortices such as vortex rings to smaller scales dominated by turbulent dissipation. The average conditional structures of Grad-CAM and Gradient-SHAP, respectively Figures \ref{fig:struct_cond_GRACAM} and \ref{fig:struct_cond_GRADIENT}, highlight regions of moderate vorticity and strain, areas known in the literature to be regions of high dissipation and turbulent kinetic energy. 

Regions with high strain are crucial in turbulence as they represent points of stretching and rotation of vortices, phenomena that drive energy transfer and mixing \cite{Hussain_1986}. Grad-CAM and Gradient-SHAP highlight exactly these areas, aligning with other studies \cite{Hussain_1986,Pickering_2020,JIMeNEZ_WRAY_1998}, which have shown that dissipation is concentrated in strain regions rather than at vorticity peaks.
Compared to SHAP, which requires predefined segmentation, Grad-CAM identifies relevant regions directly from the model gradient. This less constrained approach avoids biases introduced by segmentation and allows for the identification of physically significant regions for jet dynamics, as demonstrated by the results.

\section{Conclusion}

In this study, three different XAI algorithms were used to identify the important regions for an axisymmetric jet. Starting from time-resolved PIV data for an axisymmetric jet \citep{Roy2021}, a neural network was trained to perform a one-step prediction. To apply SHAP, it was necessary to define a criterion for segmenting the flow field in such a way as to identify regions for which the importance could be calculated using SHAP. Regions with high Shapley values indicate areas where the model focuses to make its predictions. The results show that the most relevant structures do not always correspond to those with high vorticity. This suggests that structures with moderate vorticity ($\omega_z^*$ between 0.25 and 0.5) appear to play a key role in energy transfer and jet dynamics. This behavior is consistent with previous studies that highlighted the contribution of intermediate regions to mixing and energy dissipation \citep{GEORGEwk,Crow_Champagne_1971}. Additionally, the analysis shows that the structures with the highest normalized Shapley values coincide with those that contribute most to the dissipation of turbulent kinetic energy. Regions with high dissipation not only indicate rapid loss of turbulent energy but are also closely associated with mixing and the dispersion of energy scales.

In contrast, the use of Grad-CAM and Gradient-SHAP does not require prior recognition of structures, which is advantageous as it avoids any bias. The heatmaps produced by SHAP, Grad-CAM, and Gradient-SHAP differ significantly. The main results show that: Grad-CAM identifies regions with a uniform spatial distribution, suggesting a correlation with the transport of momentum by smaller-scale vortices. The identified structures, often with a shape ratio close to 1, suggest the predominance of nearly circular vortices (e.g., vortex rings) in the jet dynamics. Vortex rings represent the dominant structures in energy transfer and turbulent mixing, supporting both jet dispersion and the formation of secondary instabilities.

\citet{Hussain_1986} introduces the concept of eduction, which refers to measuring the properties of a structure through an ensemble average of similar events. This information is useful to understand that the structures identified by XAI techniques can be seen as a form of "eduction" based on a predictive model.

\section*{DATA AVAILABILITY}
The data that support the findings of this study are available from the corresponding author upon reasonable request.

\nocite{*}
\bibliography{aipsamp}

\end{document}